\documentstyle{mn}

\newif\ifAMStwofonts

\newcommand{\etal}{{et al.}~}
\newcommand{\de}{\delta}
\newcommand{\te}{\theta}
\newcommand{\varte}{\vartheta}

\newcommand{\Sig}{\Sigma}

\newcommand{\pa}{\partial}
\newcommand{\f}{\frac}
\newcommand{\s}{\sigma}

\newcommand{\bfr}{\bmath{r}}
\newcommand{\bfs}{\bmath{s}}

\newcommand{\bfk}{\bmath{k}}
\newcommand{\bfv}{\bmath{v}}
\newcommand{\bfq}{\bmath{q}}

\newcommand{\bfp}{\bmath{p}}

\newcommand{\calF}{{\cal F}}
\newcommand{\calO}{{\cal O}}

\newcommand{\calR}{{\cal R}}

\newcommand{\calN}{{\cal N}}

\newcommand{\eps}{{\epsilon}}
\newcommand{\bc}{\begin{center}}
\newcommand{\be}{\begin{equation}}
\newcommand{\ee}{\end{equation}}
\newcommand{\ec}{\end{center}}
\newcommand{\lan}{\langle}
\newcommand{\ran}{\rangle}

\newcommand{\tte}{\tilde{\theta}}
\newcommand{\tS}{\tilde{\Sigma}}
\newcommand{\tu}{\tilde{u}}

\newcommand{\hmpc}{$h^{-1}\,{\rm Mpc}$}
\newcommand{\kms}{{${\rm km\; s^{-1}}$}}

\newcommand{\cl}{C{\L}97}

\title{Towards breaking the Omega--bias degeneracy in density--velocity 
	comparisons}

\author[M. J. Chodorowski]
     {Micha{\l} J.\ Chodorowski\\ Copernicus Astronomical Center,
      Bartycka 18, 00--716 Warsaw, Poland\\ }
\begin{document}

\maketitle

\begin{abstract} 
I derive a second-order local relation between the {\em
redshift-space\/} mass density field and the {\em real-space\/}
velocity field. This relation can be useful for comparisons between
the cosmic density and peculiar velocity fields, for a number of
reasons. First, relating the real-space velocity directly to the
redshift-space density enables one to avoid the $\Omega$-dependent
reconstruction of the density field in real space. Secondly, the
reconstruction of the three-dimensional velocity field in redshift
space, questionable because of its vorticity, is also unnecessary.
Finally, a similar relation between the {\em galaxy\/} density field
and the velocity field offers a way to break the $\Omega$-bias
degeneracy in density--velocity comparisons, when combined with an
additional measurement of the redshift-space galaxy skewness. I derive
the latter relation under the assumption of nonlinear but local bias;
accounting for stochasticity of bias is left for further study.
\end{abstract}


\section{Introduction}
\label{sec:intro}

Comparisons between the cosmic density and peculiar velocity fields of
galaxies are potentially a powerful tool to measure the cosmological
parameter $\Omega$. This is so because in the gravitational
instability paradigm the density and the velocity fields are tightly
related and the relation between them depends on $\Omega$. In the
linear regime, i.e., when the density fluctuations are significantly
smaller than unity, the density--velocity relation (DVR) is
\be
\de(\bfr) = - f^{-1}(\Omega) \nabla \cdot \bfv(\bfr) \,. 
\label{eq:i1}
\ee 
Here, $\de$ is the mass density fluctuation field, $\bfv$ is the
peculiar velocity field, $f(\Omega) \simeq \Omega^{0.6}$ and I express
distances in units of \kms. Galaxies are `test particles' which trace
the velocity field induced by the mass distribution. However, there
are both theoretical arguments (e.g., Kaiser 1984; Davis \etal 1985;
Bardeen \etal 1986; Dekel \& Silk 1986; Cen \& Ostriker 1992;
Kauffmann, Nusser \& Steinmetz 1997; Blanton \etal 1998; Dekel \&
Lahav 1998) and observational evidence (e.g., Davis \& Geller 1976;
Dressler 1980; Giovanelli, Haynes \& Chincarini 1986; Santiago \&
Strauss 1992; Loveday \etal 1996; Hermit \etal 1996; Guzzo \etal 1997;
Giavalisco \etal 1998; Tegmark \& Bromley 1998) that they are biased
tracers of the mass distribution itself. In the simplest model of
linear and local biasing, the galaxy density fluctuation field is

\be
\de_{\rm gal} = b\, \de \,, 
\label{eq:i2}
\ee
where $b$ is the linear bias factor. Then the relation between the
galaxy density and velocity field is 
\be
\de_{\rm gal}(\bfr) = - \f{b}{f(\Omega)} \nabla \cdot \bfv(\bfr) 
\,. 
\label{eq:i3}
\ee 
Thus, a comparison between the fields in question based on linear
theory yields not an estimate of $\Omega$ directly but rather of a
quantity $\beta \equiv f(\Omega) / b$. This has been called the
`$\Omega$--bias degeneracy problem'.\footnote{This terminology in the
context of linear density--velocity comparisons makes little sense to
me, since in general we can speak about degeneracy of parameters only
if we have a number of constraints on them. This is the case when
attempting to estimate $\Omega$ from the multipoles of the 
redshift-space power spectrum. Here, we have just one
constraint. Still, the terminology has become standard so I will use
it.}

To break this degeneracy in the {\it IRAS\/}--{\sc potent} comparison,
Dekel \etal (1993) used nonlinear effects. These effects in any case
should be accounted for in analyzing the data, since the observed
fields are currently smoothed over scales which are mildly
nonlinear. Dekel \etal (1993) stressed that the limits on $\Omega$
they derived `are valid only in the approximation that the linear
biasing model is correct'; simply put, they assumed linear
biasing. However, nonlinear corrections to the linear biasing
relation~(\ref{eq:i2}) are of the same order as the corrections to
the linear DVR~(\ref{eq:i1}), so in general if one accounts for
nonlinear dynamics one should also account for nonlinear biasing as
well. Such an analysis has been performed recently by Bernardeau \etal
(1999; hereafter B99), resulting in a method of disentangling $\Omega$
from (a priori) nonlinear bias.

The method of B99 assumes that both fields are given in real space,
while the observed density field, like, e.g., {\it IRAS,\/} is given
in redshift space. In density--velocity comparisons, the real-space
density field is commonly reconstructed from the redshift-space one
using an iterative technique (Yahil \etal 1991, Strauss \etal 1992),
though one-step procedures have been also invented (Fisher \etal 1995;
Taylor \& Valentine 1999). All these inversion methods aim at
correcting the positions of galaxies for their peculiar
velocities. However, to predict the velocities using an integral
version of equation~(\ref{eq:i3}) or its mildly nonlinear extension,
one has to assume the value of $\beta$. This is a serious drawback of
real-space comparisons: they necessarily assume the value of a
parameter which is to be subsequently estimated.

One might think that a way out of this problem is to perform the
comparison in redshift space. Such an analysis has been done by Nusser
\& Davis (1994), but using linear theory only. Nusser \& Davis (1994)
applied the Zel'dovich approximation to model the redshift-space DVR,
but subsequently restricted their analysis to linear regime. The
reason was, as they pointed out, that the redshift-space velocity
field is rotational at second-order, so it cannot be reconstructed
from its radial component only. Since in the present paper I aim at
disentangling $\Omega$ and bias by using nonlinear effects in the DVR,
the redshift-space analysis is also inappropriate.

A true way out of this dilemma is to relate {\em the real-space\/}
velocity field directly to the {\em redshift-space\/} density field.
Such a relation allows us to avoid at the same time problems with the
reconstruction of the three-dimensional velocity field in redshift
space and of the density field in real space. One might ask whether
this makes any sense, since the fields in question are defined in
different spaces. However, the point is that they are not: real-space
and redshift-space are useful concepts, but in fact they are merely
two different coordinate systems\footnote{Obviously, the
transformation from real space to redshift space does not conserve the
volume element and in general may be even non-invertible.}. I will
explain this in more detail later on.

In the present paper I derive a second-order relation between the
redshift space density and the real space velocity. Next I demonstrate
that, as in `real-to-real' comparisons, second-order terms in this
relation offer a method for disentangling $\Omega$ and bias. Now the
method is self-consistent because it does not make any a priori
assumption about the value of $\beta$. I give completely worked out 
an example of nonlinear but local bias. The case of stochastic bias is
left for further study.

The paper is organized as follows: in Section~\ref{sec:DVR}, I derive
the described DVR up to second-order in perturbation theory.
Specifically, in Subsection~\ref{subsec:unsmooth} I derive it in the
case of unsmoothed fields; in Subsection~\ref{subsec:smooth} I extend
it to the case of smoothed evolved fields, in a form directly
applicable to so-called density--density comparisons. Next, I compute
the expected scatter in the latter relation
(Section~\ref{sec:scatter}). In Section~\ref{sec:bias} I derive a
similar relation between the {\em galaxy\/} density field and the
velocity field under an assumption of a nonlinear but local bias. In
Section~\ref{sec:Omega_bias} I show how such a relation can be used to
break the $\Omega$--bias degeneracy in density--density
comparisons. Summary and conclusions are in Section~\ref{sec:summ}.

\section{Redshift-space density versus real-space velocity}
\label{sec:DVR}
The purpose here is to derive a second-order local relation between
the redshift-space density field and the real-space peculiar velocity
field. Intrinsic motions of galaxies along the line of sight cause
distortion in galaxy redshift surveys. On one hand, this distortion
complicates measurements of the statistical properties of the large
scale galaxy distribution and also cosmic density--velocity
comparisons. On the other hand, it contains valuable information on
the structure of the peculiar velocity fields. Since the amplitude of
peculiar velocities depends on $\Omega$, measurement of the degree of
the distortion can serve as a method of measuring this parameter (see
Hamilton 1997: a review on linear redshift distortions). Similarly to
density--velocity comparisons, linear studies of redshift distortions
yield merely an estimate of $\beta$ and in order to disentangle
$\Omega$ from bias, nonlinear effects have been recently investigated
(Taylor \& Hamilton 1996; Heavens, Matarrese \& Verde 1998;
Scoccimarro \etal 1998).

On large scales ($\ga$ several \hmpc), coherent flows lead to
compression of structures along the line of sight. This compression
reaches maximum at a turnaround radius, where the structure appears
totally collapsed along the line of sight and thereafter begins to
invert itself. On small scales ($\la$ a few \hmpc), high velocities of
galaxies in clusters stretch out the structure in redshift space,
creating the so-called `finger-of-God effects'. These effects are
commonly interpreted as an evidence for virialisation of
clusters. However, this is not necessarily the case. During
three-dimensional highly-nonlinear gravitational collapse (but still
before shell crossing), a structure attains infall velocities which
are typically much bigger compared to its size multiplied by the
Hubble constant. Therefore, young unvirialized clusters, still partly
in an infall phase, may also appear in redshift space as
`fingers-of-God' (Chodorowski 1990).

A perturbative approach which I will adopt here is valid for scales
bigger than turnaround. Then the `compression effect' dominates and as
we will see below, the redshift-space overdensity is increased
compared to the corresponding real-space overdensity.

\subsection{Unsmoothed fields}
\label{subsec:unsmooth}
At the beginning my analysis will be similar to that of Heavens,
Matarrese \& Verde (1998), who studied nonlinear corrections to the
redshift-space distortion of the power spectrum. Let $\bfr$ and $\bfs$
be the coordinates in real and redshift space, respectively, with the
Local Group at the origin. The redshift-space position of a galaxy is
related to its real-space position by (Kaiser 1987)
\be 
\bfs = \bfr \left[1 + \f{u_{}(\bfr)}{r} \right] .
\label{eq:realtored}
\ee 

Here, $u_{}(\bfr) \equiv \bfv \cdot \bfr / r$ and velocities are
expressed relative to the Local Group. Let $\rho_{}$ and $\rho_{s}$
be the density fields in real and redshift space. The mean density may
vary in space, due to selection effects; I will denote the mean
densities as $\phi_{}(\bfr)$ and $\phi_{s}(\bfs)$. The density
contrast in redshift space is then defined by $1 + \de_{s}(\bfs) =
\rho_{s}(\bfs)/\phi_{s}(\bfs)$; similarly for the real-space
density contrast.

Number conservation of galaxies implies $\rho_{s}(\bfs) {\rm d}^3 \bfs =
\rho_{}(\bfr) {\rm d}^3 \bfr$. This yields
\begin{eqnarray}
\left[1 + \de_{s}(\bfs)\right] \phi_{s}(\bfs) 
\!\!\!\!\! &=& \!\!\!\!\!\! \left[1 + \f{u_{}(\bfr)}{r} \right]^{-2}
\left[1 +  u_{}'(\bfr) \right]^{-1} \times 
\nonumber \\
\!\!\!\!\! &~& \!\!\!\!\! \left[1 + \de_{}(\bfr)\right] \phi_{}(\bfr)
\,,
\label{eq:numb_cons}
\end{eqnarray} 
where $' \equiv \partial/\partial r$. I will make the standard
distant-observer approximation: that the survey is large enough so
that the modes with wavenumbers $k \la r^{-1}$ are negligible. Then
in the above expression we can neglect the term $u_{}(\bfr)/r$;
furthermore, a Taylor expansion of $\phi(s)$ yields $\phi(r)$ plus a
negligible correction. Second-order expansion of the above equation is
then

\be
\de_{s}(\bfs) = \de_{}(\bfr) - u_{}'(\bfr)
- \de_{}(\bfr) u_{}'(\bfr) + \left[u_{}'(\bfr)\right]^2
\,.
\label{eq:sr_rel}
\ee 

I will now use the fact that in real space the density field is up to
second order a local function of the two velocity scalars: the
expansion (the velocity divergence), $\te$, and the shear, $\Sig$.
Specifically (Chodorowski 1997; cf.~Gramann 1993, Catelan \etal 1995,
Mancinelli \& Yahil 1995)

\be
\de_{}(\bfr) = - f^{-1} \te(\bfr) +
{\textstyle \f{4}{21}} f^{-2} \te^2(\bfr) 
- {\textstyle \f{2}{7}} f^{-2} \Sig^2(\bfr) \,.
\label{eq:c97} 
\ee
Here, 
\be
\Sig^2 \equiv \Sig_{ij} \Sig_{ij} \,,
\label{eq:shear_scalar} 
\ee
\be
\Sig_{ij} \equiv {\textstyle \f{1}{2}} 
\left(\pa v_i/\pa r_j + \pa v_j/\pa r_i \right)
- {\textstyle \f{1}{3}} \de^{K}_{ij} \te
\label{eq:shear_tensor} 
\,,
\ee
\be
\te \equiv \nabla_{\bfr} \cdot \bfv
\,,
\label{eq:div} 
\ee 
and I have neglected the weak $\Omega$-dependence. The symbol
$\de^{K}_{ij}$ denotes the Kronecker delta. Using
expression~(\ref{eq:c97}) in~(\ref{eq:sr_rel}) gives up to second
order
\begin{eqnarray}
\de_{s}(\bfs) 
\!\!\!\!\! &=& \!\!\!\!\! - f^{-1} \te(\bfr) - u_{}'(\bfr)
+ f^{-1} \te(\bfr) u_{}'(\bfr) + \left[u_{}'(\bfr)\right]^2
+ \nonumber \\
\!\!\!\!\! &~& \!\!\!\!\!
{\textstyle \f{4}{21}} f^{-2} 
\left[\te^2(\bfr) - {\textstyle \f{3}{2}} \Sig^2(\bfr)\right]
\,.
\label{eq:sr_dv}
\end{eqnarray} 

Thus, the redshift-space density field at a position $\bfs$ is a
function of the derivatives of the real-space velocity field at a
corresponding position $\bfr$. This is in fact a {\em local\/}
relation between the redshift-space density and the real-space
velocity. Real-space and redshift-space are useful notions, but
actually they are merely two different mappings of the same physical
space to $\calR^3$. The coordinate $\bfr$ and the corresponding
coordinate $\bfs$, given by equation~(\ref{eq:realtored}), label the
same point in physical space. Therefore, any local relation between
the redshift-space quantity-one and the real-space quantity-two
naturally expresses quantity-one at a position $\bfs$ in terms of
quantity-two at a position $\bfr$. However, this is inconvenient,
because in the process of comparison, for each position $\bfr$ we
should, given the velocity field, seek the corresponding
$\bfs$. Fortunately, the value of any real-space function $g$ at
$\bfr$ can be expressed as its value at $\check{\bfr} \equiv \bfs$
plus Taylor-expansion corrections, which will depend on the
velocity. To second order we have

\be
g(\bfr) = g(\bfs) - u_{}(\bfs) g'(\bfs) 
\,.
\label{eq:rtos}
\ee 
This allows us to cast the RHS side of Equation~(\ref{eq:sr_dv}),
involving the velocity evaluated at $\bfr$, to another form, but also
requiring only the velocity, evaluated at $\check{\bfr} \equiv
\bfs$. (I have used a `check' mark to emphasize that $\check{\bfr}$ is
a real-space point but in general different from $\bfr$, which is
related to $\bfs$ by eq.~\ref{eq:realtored}). The result is the
relation between the redshift-space density and the real-space
velocity, all evaluated at $\bfs$:

\be
\de_{s} =  - f^{-1} \te - u_{}' 
+ \left[u_{}(f^{-1} \te + u_{}')\right]'
+ {\textstyle \f{4}{21}} f^{-2}
\left(\te^2 - {\textstyle \f{3}{2}} \Sig^2\right) .
\label{eq:ss_dv}
\ee 

What does it exactly mean? It means that if we select a point $\bfs$
in redshift space, to estimate the redshift-space density contrast at
this point we need to evaluate the RHS of the above equation at a
real-space point $\check{\bfr} = \bfs$. Conversely, {\em if we select
first a point $\bfr$ in real space, evaluating the RHS of the above
equation at this point we estimate the redshift-space density contrast
at a redshift-space point $\check{\bfs} = \bfr$.} In other words, we
can now treat the two fields as if they were given in the same
coordinate system: in fact we are now comparing the density with the
velocity at two (slightly) different physical points (since
$\check{\bfs} \ne \bfs$), but the Taylor terms in the resulting DVR
provide the necessary correction. 

I reiterate that equation~(\ref{eq:ss_dv}) involves real-space
derivatives of the real-space velocity field. Having measured the
peculiar velocity of a galaxy, we have a choice: we can assign the
velocity either to the galaxy's distance, or to its redshift. The
former procedure leads to construction of the velocity field in real
space; the latter to construction of the velocity field in redshift
space. Thus, the definition of the redshift-space velocity field,
$\bfv_{\bfs}$, is $\bfv_{\bfs}(\bfs) \equiv \bfv(\bfr)$. This field is
{\em not\/} vorticity-free: $\nabla_{\bfs}\times \bfv_{\bfs}(\bfs) =
\nabla_{\bfs} \times \bfv(\bfr) \ne \nabla_{\bfr} \times \bfv(\bfr)$,
the last quantity being the vorticity of the velocity field in real
space, which vanishes. On the other hand, the difference between
$\nabla_{\bfs}\times \bfv_{\bfs}(\bfs)$ and $\nabla_{\bfr} \times
\bfv(\bfr)$, and hence the resulting vorticity in redshift space, is
already of second-order in velocity. In sum, while the linear velocity
field in redshift space is irrotational the nonlinear one is not and
therefore cannot be reconstructed from its radial component only.

As stated earlier, in density--velocity comparisons a common procedure
is to first reconstruct the real-space density from the redshift-space
one. Relation~(\ref{eq:ss_dv}) provides a basis to avoid this step.
The application of this relation to density--density comparisons is
obvious: (a) take the appropriate derivatives of the real-space
velocity field at $\bfr$; (b) compare the result directly to the
redshift-space density contrast at $\check{\bfs} = \bfr$. As for
velocity--velocity comparisons, relation~(\ref{eq:ss_dv}) can be
straightforwardly used to reconstruct the real-space velocity field
directly from the redshift-space density field. The real-space
velocity field is irrotational thus describable as a gradient of some
potential. Relation~(\ref{eq:ss_dv}) reduces then to a differential
equation (in real space) for the velocity potential, with the source
term $S(\bfr) \equiv \de_{s}(\check{\bfs} = \bfr)$. This equation can
be solved perturbatively: at linear order, by the method of Nusser \&
Davis (1994);\footnote{The linear part of equation~(\ref{eq:ss_dv})
coincides with equation~(6) of Nusser \& Davis (1994) in the distant
observer limit.} at second order similarly, with the source term
modified by the nonlinear terms in velocity, approximated by linear
solutions. In the present paper I concentrate on the application of
the relation~(\ref{eq:ss_dv}) to density--density comparisons; the
application to velocity--velocity comparisons will be addressed
elsewhere.

The second-order DVR~(\ref{eq:ss_dv}) has been derived in a quite 
simple and natural way. There is, however, a more formal way of
deriving it and, more importantly, generalizing for {\em all\/}
orders. For convenience, introduce a {\em scaled\/} velocity field,

\be
\tilde{\bfv} = - f^{-1} \bfv 
\,.
\label{eq:scaled_vel}
\ee 
Equation~(\ref{eq:ss_dv}) then reads 

\be
\de_{s}|_{\check{\bfs} = \bfr} = \left. \left\{ \tte + f \tu_{}' 
+ f \left[\tu_{}(\tte + f \tu_{}')\right]'
+ {\textstyle \f{4}{21}}
\left(\tte^2 - {\textstyle \f{3}{2}} \tS^2\right) \right\} \right|_{\bfr}
\!\!,
\label{eq:scaled_dv}
\ee 
where for any function $g$, $\tilde{g} \equiv - f^{-1} g$. Applying
the approach of Scoccimarro \etal (1998), in Appendix~\ref{app:scocci}
I show that in general

\be
\de_{s}|_{\check{\bfs} = \bfr} = 
\sum_{n = 0}^{\infty} \f{1}{n!} f^n
\left. \left[ \tu_{}^n (\de + f \tu_{}') \right]^{(n)} \right|_{\bfr} .
\label{eq:scocci}
\ee 
Here, the superscript `$^{(n)}$' denotes `$\partial^n/\partial
z^n$'. This equation will yield the desired DVR when we substitute for
the real-space density contrast $\de(\bfr)$ its local estimate from
velocity; in other words, when we make use of the {\em real-space\/}
DVR. Equation~(\ref{eq:scaled_dv}) is easily obtained
from~(\ref{eq:scocci}) when the second-order relation~(\ref{eq:c97})
is used and the series above is consequently truncated at second-order
($n=1$) terms. However, extensions of real-space DVRs beyond second
order have been studied both analytically and numerically (Nusser
\etal 1991; Mancinelli \etal 1994; Chodorowski \& {\L}okas 1997,
hereafter \cl; Chodorowski \etal 1998; B99; Ganon \etal 1999),
resulting in a robust analytical description of the DVR in the mildly
nonlinear regime.\footnote{I.e., for density fluctuations of order
a few.}  Specifically, B99 and Ganon \etal (1999) found nonlinear
formulas for density in terms of the velocity derivatives, which have
proven to be very accurate, when tested against N-body
simulations. One might thus think that using them in
equation~(\ref{eq:scocci}) would yield a very good description of the
mildly nonlinear local relation between the redshift-space density and
the real-space velocity.

In fact, this is not so simple and in the present paper I restrict the
analysis to second order. The reason is that
equations~(\ref{eq:scocci}) or~(\ref{eq:scaled_dv}) are not directly
applicable to real data, since the observed density and velocity
fields are always smoothed over some scale. As in real space, the
analog of equation~(\ref{eq:scaled_dv}) -- or of its higher-order
extension -- for smoothed fields will significantly differ from its
unsmoothed counterpart (see the next subsection). Therefore, smoothing
{\em should\/} be accounted for. This, however, turns out to be a
non-trivial task, even in real space (\cl, Chodorowski \etal 1998,
B99), and calculations in redshift space are even more difficult.
Restricting the analysis to second order will enable me to
straightforwardly calculate the effects of smoothing.  Furthermore, as
I will show it later, a second-order analysis is sufficient for the
purpose of the present paper, i.e., outlying an idea of breaking the
degeneracy between $\Omega$ and bias in density--velocity comparisons
by nonlinear effects. Therefore, including orders higher than second
would make the analysis more complicated than is necessary here. 

At the end of this subsection, let us compare contributions of the
linear and nonlinear terms in relation~(\ref{eq:ss_dv}). To simplify
the comparison, consider a spherical top-hat overdensity.  Then
$\Sig^2 = 0$, $\te = \rm const$, $u_{}' = {\textstyle \f{1}{3}} \te$,
and hence

\be
\de_{s} =  - \left(1 + {\textstyle \f{1}{3}} f\right) f^{-1} \te 
+ {\textstyle \f{4}{21}} \left(1 + {\textstyle \f{7}{4}} f + 
{\textstyle \f{7}{12}} f^2 \right) f^{-2} \te^2
\,.
\label{eq:tophat}
\ee 
The real-space counterpart of the above `redshift-to-real' relation
can be obtained by setting the $f(\Omega)$ factors in the parentheses
equal to zero, yielding correctly a special case of
equation~(\ref{eq:c97}):

\be
\de =  - f^{-1} \te + {\textstyle \f{4}{21}} f^{-2} \te^2
\,.
\label{eq:rr_tophat}
\ee 
The amplitude of the second-order term relative to the linear one in
this real-space relation is given for $\te = 1$ by $(4/21) f^{-1}
\simeq 0.2 f^{-1}$. Hence, while for a low-$\Omega$ Universe
real-space nonlinear effects can be significant, for a high-$\Omega$
($\Omega \sim 1$) Universe they are rather weak (Dekel \etal
1993). However, nonlinear effects in the relation~(\ref{eq:tophat})
are significantly stronger. Peculiar velocities increase the
redshift-space overdensity at both orders, but while the linear term
is increased only by $33 \%$ for $\Omega = 1$, the second-order term
is increased by more than a factor of three. Thus, the relative
importance of the nonlinear terms in the `redshift-to-real' DVR is
significantly greater than of those in the real-space DVR, and so is
the sensitivity of the degeneracy-breaking method presented here.

\subsection{Smoothed fields}
\label{subsec:smooth}

As stated earlier, the density and the velocity fields reconstructed
from observations are always smoothed over some (large) scale $R$. At
linear order, the DVR for smoothed evolved fields follows immediately
from its counterpart for unsmoothed fields. From the linear part of
equation~(\ref{eq:scaled_dv}) and equation~(\ref{eq:div}) we have
$\de_{s}^R = \nabla_{\bfr} \cdot (\tilde{\bfv}^R) + f (\tu^{R})'$,
because the operations of smoothing ($^R$) and differentiation
commute. However, the operations of smoothing and multiplication do
not and the derivation of an analog of equation~(\ref{eq:scaled_dv})
for smoothed fields including second-order terms is already a
non-trivial task. Here I will derive the relation in a form applicable
to so-called density--density comparisons, i.e., expressing density in
terms of velocity. In such comparisons, a large-scale velocity field
is used to reconstruct the associated mass density field, which is
subsequently compared to the galaxian density field. To simplify the
notation, I will from now on implicitly assume that all fields are
smoothed at scale $R$ and will drop the superscript $^R$ on all
variables.

Being a non-local operation, smoothing inevitably introduces some
scatter in any local relation between nonlinear variables. As an
estimator of the smoothed density from the smoothed velocity we cannot
therefore do better than to adopt the conditional mean, the mean
redshift-space density contrast {\em given\/} in general a set of
velocity derivatives, $\lan \de_{s} \ran \vert_{\pa \tilde{v}_i/\pa
r_j, \, \pa^2 \tilde{v}_i/\pa r_j \pa r_k, \, \ldots}$. Second-order
relation~(\ref{eq:scaled_dv}) for unsmoothed fields involves velocity
derivatives up to second order, namely $\tte$, $\tS$, $\tu'$ (first
order) and $\tte'$, $\tu''$ (second). In principle, it would be then
natural to compute the quantity $\lan \de_{s} \ran \vert_{\tte, \tS,
\tu', \tte', \tu''}$. In practice, however, this is not the best thing
to do. (Leaving aside extreme complexity of such a calculation!)  The
differentiation of the noisy peculiar velocity field should be done
with great care and even so it still leads to dangerous biases (Dekel
\etal 1998). The current incarnation of the {\sc potent} algorithm, as
an estimator of the real-space mass density from the velocity field
uses a nonlinear formula but involving only first-order velocity
derivatives (Dekel \etal 1998, Sigad \etal 1998). Second-order
derivatives of the velocity fields reconstructed from current catalogs
of peculiar velocities probably cannot be reliably estimated at
all. It would be thus more reasonable to compute the quantity $\lan
\de_{s} \ran \vert_{\tte, \tS, \tu'}$.\footnote{Chodorowski~1997
derived a related expression for the {\em real-space\/} density, $\lan
\de \ran \vert_{\tte, \tS}$.}

However, a simple method of breaking the $\Omega$--bias degeneracy in
density--density comparisons, which I will present later on, relies on
a relation between $\de_{s}$ and just $\tte$. In such a relation,
unlike in relation~(\ref{eq:scaled_dv}), both the (total) linear term
and the second-order term will be a separable function of $\Omega$ and
the velocity derivatives.\footnote{Obviously, this is always the case
when the velocity field is described by one variable.} (See the
simplified relation~\ref{eq:tophat}). This will enable one to perform
the comparison in a model-independent way, as described below.
Therefore, as the only descriptor of the velocity field I finally
choose the variable $\tte$. An estimator of the redshift-space density
constructed exclusively from $\tte$ has clearly bigger variance than
an estimator constructed jointly from $\tte$, $\tS$ and $\tu'$. Still,
this variance can be sufficiently reduced if the velocity field is
sampled at a large-enough number of independent volumes.

Taking the conditional mean of both sides of
equation~(\ref{eq:scaled_dv}) we have

\be
\lan \de_{s} \ran \vert_{\tte} = f \lan \tu_{}' \ran|_{\tte}
+ \lan \tte + \Delta_2 \ran|_{\tte}
\,,
\label{eq:mean_dv}
\ee
where

\be
\Delta_2 \equiv f \left[\tu_{}(\tte + f \tu_{}')\right]'
+ {\textstyle \f{4}{21}}
\left(\tte^2 - {\textstyle \f{3}{2}} \tS^2\right)
\,.
\label{eq:Delta}
\ee

Define now the variable 

\be
\gamma \equiv \tte + \Delta_2
\,.
\label{eq:beta}
\ee 
\cl\ derived a general expression for the conditional mean for two 
mildly nonlinear smoothed variables, given Gaussian initial
conditions, assumed also here. An additional assumption was that the
variables are equal at linear order. Expanding $\gamma$ to second
order we have $\gamma = \tte_1 + \tte_2 + \Delta_2$, where $\tte_i$
are $i$-th order perturbative contributions to $\tte$; in general
$\tte_i$ is $\calO(\tte_1^i)$. One can see that $\Delta_2$, like
$\tte_2$, is already $\calO(\tte_1^2)$. Thus, the variables $\gamma$
and $\tte$ are indeed equal at linear order. To second order for such
variables (\cl),

\be
\lan \gamma \ran |_{\tte} = \tte 
+ a_2 \left(\tte^2 - \lan \tte^2 \ran \right)
\,.
\label{eq:be_te}
\ee
Here, $a_2 \equiv (S_{3 \gamma} - S_{3 \tte})/6$ and $S_{3 \gamma}$ ($S_{3
\tte}$) denotes the skewness of the variable $\gamma$ ($\tte$). Using
the definition~(\ref{eq:beta}) of $\gamma$, at leading order we obtain:

\be
a_2 = {\textstyle \f{1}{2}} \left\lan \tte_1^2 \right\ran^{-2} 
\left\lan \tte_1^2 \Delta_2 \right\ran  
\,.
\label{eq:a_2}
\ee

By isotropy we have 

\be
\lan \tu_{}' \ran|_{\tte} = {\textstyle \f{1}{3}} \tte 
\,.
\label{eq:u'_mean}
\ee

This, together with equations~(\ref{eq:beta}) and~(\ref{eq:be_te}),
used in equation~(\ref{eq:mean_dv}) yields

\be
\lan \de_{s} \ran \vert_{\tte} = 
\left(1 + {\textstyle \f{1}{3}} f \right) \tte
+ a_2 \left(\tte^2 - \lan \tte^2 \ran \right)
\,,
\label{eq:mean_dv2}
\ee
or, transforming back to the plain velocity divergence,  

\be
\lan \de_{s} \ran \vert_{\te} = 
- \left(1 + {\textstyle \f{1}{3}} f \right) f^{-1} \te
+ a_2 f^{-2} \left(\te^2 - \bigl\lan \te^2 \bigr\ran \right)
\,.
\label{eq:mean_dv3}
\ee
This is a second-order relation between the redshift-space mass
density, at $\check{\bfs} = \bfr$, and the real-space velocity
divergence, at $\bfr$. The calculation of the coefficient $a_2$ for
different smoothing windows is outlined in
Appendix~\ref{app:nobias}. In analyzes of observational data the
fields are commonly smoothed with a Gaussian window. The value of
$a_2$ for a Gaussian smoothing function is given by
equation~(\ref{app:a_2G}).

The coefficient $a_2$ is a measure of the degree of nonlinearity in
the DVR. Since a method for breaking the $\Omega$--bias degeneracy,
which I will present in Section~\ref{sec:Omega_bias}, relies on the
value of $a_2$, we should compute it accurately. Smoothing clearly
affects the value of $a_2$ (see Appendix~\ref{app:nobias}). For
simplicity, consider the case of a top-hat smoothing. From
equation~(\ref{app:a_2TH}) it follows that for $\Omega = 1$, the value
of $a_2$ for unsmoothed fields, $a_2^{(\rm un)}$, overestimates the
true value by almost $40$\% for the spectral index $n = -1$ and by
more for bigger values of $n$. Thus, smoothing changes the value of
the coefficient $a_2$ significantly so it was indeed necessary to
account for it. On the other hand, it is worth noting that the true
value for $\Omega = 1$ and $n = -1$ is still more than a factor of two
bigger than its real-space counterpart,
equation~(\ref{app:a_2THreal}).

\section{Scatter in the relation}
\label{sec:scatter}

A local relation between the density and the velocity divergence
derived in previous Section has a scatter, present already at linear
order. The r.m.s. value of the scatter is given by the square root of
the conditional variance. At linear order we have
\begin{eqnarray}
\left.\left\lan \left(\de_{s} - \lan \de_{s} \ran|_{\tte} \right)^2 
\right\ran \right|_{\tte} 
\!\! &=& \!\! 
f^2 \left.\left\lan\left(\tu' 
- {\textstyle \f{1}{3}} \tte \right)^2 \right\ran \right|_{\tte} 
\nonumber \\
 \!\! &=& \!\!
{\textstyle \f{1}{9}} f^2 \tte^2 
- {\textstyle \f{2}{3}} f^2 \tte \lan \tu' \ran|_{\tte} 
+ f^2 \lan \tu'^2 \ran|_{\tte}
\,.
\label{eq:scatt_in}
\end{eqnarray}
The quantity $\lan \tu_{}' \ran|_{\tte}$ is given by
equation~(\ref{eq:u'_mean}). The calculation of the quantity $\lan
\tu'^2 \ran|_{\tte}$ is presented in Appendix~\ref{app:u'}. The result
is given by equation~(\ref{appb:fin}), where $\s_{\tte}^2$ is the
variance of the field $\tte$. This yields  
$\left.\left\lan \left(\de_{s} - \lan \de_{s} \ran|_{\tte} \right)^2 
\right\ran \right|_{\tte} = {\textstyle \f{4}{45}} f^2 \s_{\tte}^2$, or

\be
\left.\left\lan \left(\de_{s} - \lan \de_{s} \ran|_{\tte} \right)^2 
\right\ran \right|_{\tte}^{1/2} = {\textstyle \f{2}{3 \sqrt{5}}} f
\s_{\tte} 
\,.
\label{eq:scatt_fin2}
\ee 
The scatter is of the order of the redshift-space correction to
the linear term in equation~(\ref{eq:mean_dv2}). When compared to the
total linear term in equation~(\ref{eq:mean_dv2}), the scatter is
relatively small: for $\tte \sim \s_{\tte}$, it is $\sim 20\%$ for
$\Omega = 1$ and accordingly less for smaller $\Omega$. This implies
that redshift-space density is well (though not as well as real-space
density) correlated with real-space velocity-divergence.

\section{Galaxian density versus velocity}
\label{sec:bias}
As already stated in the introduction, it is currently clear that
galaxies are biased tracers of the mass distribution. In this section
I will derive a relation between the redshift-space {\em galaxy\/}
density field and the real-space velocity field under an assumption of
a nonlinear but local bias. This is only a toy model for bias because
there are good reasons to believe that bias is in fact somewhat
stochastic (Dekel \& Lahav 1998; Pen 1998; Tegmark \& Peebles 1998;
Blanton \etal 1998; Tegmark \& Bromley 1998). However, a number of
important conclusions can be drawn already from this model.

Equation~(\ref{eq:scocci}) has been derived from conservation of
galaxy numbers, so the densities appearing in it are in fact the {\em
galaxy\/} densities $\de_{s}^{\rm (g)}$ and $\de_{s}^{\rm (g)}$
(redshift- and real-space galaxy density contrasts respectively).
Hence, a more general form of equation~(\ref{eq:scocci}) is

\be
\left. \de_{s}^{\rm (g)}\right|_{\check{\bfs} = \bfr} = 
\sum_{n = 0}^{\infty} 
\f{1}{n!} f^n \left. \left[ \tu_{}^n (\de^{\rm (g)} + 
f \tu_{}') \right]^{(n)} \right|_{\bfr} \,;
\label{eq:scocci_gal}
\ee 
the form~(\ref{eq:scocci}) implicitly assumes no bias between the
distribution of galaxies and mass. In a local bias model, the
real-space galaxy density contrast is assumed to be in general a
nonlinear, but local function of the mass density contrast (Fry \&
Gazta\~naga 1993; see also Juszkiewicz \etal 1995),

\be
\de^{\rm (g)}(\bfr) = \calN\left[\de(\bfr)\right]
\,. 
\label{eq:bias}
\ee
Given a (nonlinear) local formula for the real-space mass density in
terms of the real-space (first-order) velocity derivatives, $\de(\bfr)
= \calF\left[\pa v_i/\pa r_j(\bfr)\right]$,
equation~(\ref{eq:scocci_gal}) yields

\be 
\left.\de_{s}^{\rm (g)}\right|_{\check{\bfs} = \bfr} = \sum_{n = 0}^{\infty}
 \f{1}{n!} f^n \left. \left\{ \tu_{}^n \left[\calN \circ 
\calF(\pa v_i/\pa r_j) + f \tu_{}'\right] \right\}^{(n)}
\right|_{\bfr} 
\!.
\label{eq:dv_bias}
\ee 

Up to second order, $\calF$ is given by expression~(\ref{eq:c97}) and

\be
\calN\left[\de(\bfr)\right] = b \de(\bfr) + {\textstyle \f{1}{2}} b_2 
\left(\de^2(\bfr) - \lan \de^2 \ran \right)  
\,.
\label{eq:bias_2}
\ee 
Here, $b$ and $b_2$ are respectively the linear and nonlinear
(second-order) bias parameters. The term $\lan \de^2 \ran$
ensures that the mean value of $\de^{\rm (g)}$ vanishes, as
required. Using this in equation~(\ref{eq:dv_bias}) and truncating
the series at second-order terms we obtain
\begin{eqnarray}
\left.\de_{s}^{\rm (g)}\right|_{\check{\bfs} = \bfr}
\!\!\!\!\! &=& \!\!\!\!\!
\left\{ b \tte + f \tu_{}' + f \left[\tu(b \tte + f \tu')\right]' 
\right.
\nonumber \\
\!\!\!\!\! &~& \!\!\!
\left. \left. + {\textstyle \f{4}{21}} b \left(\tte^2 - 
{\textstyle \f{3}{2}} \tS^2\right) 
+ {\textstyle \f{1}{2}} b_2
\left(\tte^2 - \lan \tte^2 \ran \right) \right\} \right|_{\bfr}
.
\label{eq:dv_bias2}
\end{eqnarray} 

This is a second-order local relation between the redshift-space
galaxian density and the real-space velocity, derived under the
assumption of local bias. This relation is for unsmoothed fields. To
account for the effects of smoothing and to enable a model-independent
density--velocity comparison we proceed analogously to when
constructing the corresponding relation for the redshift-space {\em
mass\/} density. Namely, we compute the expectation value of the
redshift-space galaxian density given the velocity divergence. The
result is

\be
\lan \de_{s}^{\rm (g)} \ran \vert_{\tte} = 
\left(b + {\textstyle \f{1}{3}} f \right) \tte
+ a_2^{\rm (g)} \left(\tte^2 - \lan \tte^2 \ran \right)
\,,
\label{eq:dv2_bias}
\ee
or, transforming back to the plain velocity divergence,  

\be
\lan \de_{s}^{\rm (g)} \ran \vert_{\te} = 
- \left(b + {\textstyle \f{1}{3}} f \right) f^{-1} \te
+ a_2^{\rm (g)} f^{-2} \left(\te^2 - \lan \te^2 \ran \right)
\,.
\label{eq:dv2_bias_plain}
\ee
The calculation of the coefficient $a_2^{\rm (g)}$ is outlined in
Appendix~\ref{app:bias}. The result for Gaussian-smoothed fields is
given by formula~(\ref{app:a_2G_g}). The coefficient $a_2^{\rm (g)}$
depends on three parameters: $\Omega$, $b$ and $b_2$. 

In section~\ref{sec:DVR} I argued that in the related {\em
mass}-density versus velocity-divergence relation the nonlinear
corrective term has fairly big amplitude. However, in the above
expression for the galaxy density it may even vanish, if in
equation~(\ref{app:a_2G_g}) for the coefficient $a_2^{\rm (g)}$ the
nonlinear biasing term of negative sign cancels exactly the terms due
to nonlinear dynamics. Negative nonlinear biasing of the {\it IRAS\/}
density field has been inferred from a preliminary analysis of its
bispectrum (Scoccimarro, private communication). This makes physical
sense, since the {\it IRAS\/} galaxies are underrepresented in the
cores of rich galaxy clusters, which is effectively a nonlinear {\em
antibiasing\/} operation on the mass density field. This may partly
explain why in the {\it IRAS\/}--{\sc potent} comparison the measured
nonlinear effects are weak (Dekel \etal 1993), although in real-space
analyses they are generally weaker anyway.

\section{Breaking the $\bmath{\lowercase{\Omega}}$--bias degeneracy}
\label{sec:Omega_bias}

Second-order equation~(\ref{eq:dv2_bias_plain}) can be used to
estimate $\Omega$ separately from bias in density--density
comparisons. As first pointed out by Dekel \etal (1993), the key point
of the idea of breaking the $\Omega$--bias degeneracy by nonlinear
effects is not to correct for them in the density reconstruction, but
rather to use them as additional information. The degeneracy-breaking
method presented here is similar to that proposed by B99. However, the
method is now self-consistent, because it no longer needs to make any
a priori assumptions about the values of estimated parameters. To
apply the method of B99, we should first reconstruct the {\em
nonlinear\/} real-space density field. In order to do this, we would
have to assume simultaneously the values of $\Omega$, linear bias and
nonlinear (quadratic) bias. Here, we relate the real-space velocity
field directly to the redshift-space density field and this problem
disappears.

First steps of the degeneracy-breaking analysis can be outlined as
follows: the output of a {\sc potent}-like reconstruction machinery
should be real-space plain velocity divergence ($\te$), a quantity
derived directly from the data. Next, the velocity divergence should
be compared to the redshift-space galaxian density, also a
model-independent quantity. Since $\Omega$ and bias factors appear
only in the coefficients of the density versus velocity-divergence
relation (eq.~\ref{eq:dv2_bias_plain}), this comparison is clearly
model-independent. Specifically, the redshift-space galaxian density
should be plotted against the velocity divergence and a second-order
polynomial should be fitted to the (hopefully) apparent
correlation. The fitted linear and quadratic coefficients give us two
equations for three variables $\Omega$, $b$ and $b_2$. We need
therefore an additional constraint on these variables.

As this constraint we can adopt the large-scale galaxian density
skewness, since it is a second order effect, involving $b$ and
$b_2$. Galaxian skewness can be measured only in redshift
space. Therefore, we need a theoretical relation between the
redshift-space galaxy density skewness $S_{3 s}^{\rm (g)}$ (which we
can measure) and the redshift-space mass density skewness $S_{3 s}$
(which we can compute). Such a relation will depend on the real-space
bias factors $b$ and $b_2$. This dependence may be quite complicated,
since the operations of bias and redshift-space mapping do not
commute.

The issue of non-commutivity of bias and redshift-space mapping was
addressed in some detail by Scoccimarro, Couchman \& Frieman
(1998). Specifically, they compared the redshift-space hierarchical
amplitude for the galaxy field, $Q_s^{\rm (g)}$, to that predicted if
bias and the mapping commuted. (The hierarchical amplitude is defined
as the ratio of the bispectrum to the relevant sum of products of the
power spectra.) In the local bias model, the {\em real-space\/}
galaxian hierarchical amplitude, $Q^{\rm (g)}$, is related to the
real-space mass hierarchical amplitude, $Q$, by $Q^{\rm (g)} = Q/b +
b_2/b^2$ (Fry 1994). If the assumption of commutivity were correct,
this would imply

\be
Q_s^{\rm (g)} = Q_s/b + b_2/b^2 \,,
\label{eq:Q_s}
\ee
where $Q_s$ is the redshift-space mass hierarchical amplitude. 

Scoccimarro \etal (1998) showed that the actual $Q_s^{\rm (g)}$
differs from $Q_s/b + b_2/b^2$. However, the differences are quite
small: at most (for some values of angle) $20$\% for $\Omega = 1$ and
respectively smaller for smaller $\Omega$. Moreover, as a function of
angle, they fluctuate in sign. The galaxy skewness is essentially an
integral over the galaxy hierarchical amplitude. Therefore, the
differences between the actual skewness and its approximate value from
an analog of equation~(\ref{eq:Q_s}) will largely cancel out. Using
equation~(\ref{eq:Q_s}), a simple calculation yields

\be
S_{3 s}^{\rm (g)} = \f{S_{3 s}}{b} + 3 \f{b_2}{b^2} 
\,.
\label{eq:S_3}
\ee
For the reasons stated above, we can expect this equation to be accurate to
$\sim 10$\% or less.

It should be emphasized that {\em approximate commutivity applies only
to particular combinations of some statistical quantities and not to
the quantities alone.} In the case of skewness, this combination is
the ratio of the third moment to the square of the second. It is
straightforward to show that at leading order, the mass second moment
is $\lan \de_{s}^2 \ran = (1 + {\textstyle \f{2}{3}} f + {\textstyle
\f{1}{5}} f^2) \lan \de^2 \ran$ and the galaxy second moment is $\lan
[\de_{s}^{\rm (g)}]^2 \ran = b^2 (1 + {\textstyle \f{2}{3}} \beta +
{\textstyle \f{1}{5}} \beta^2) \lan \de^2 \ran$, where $\beta \equiv
f(\Omega)/b$. Therefore, $\lan [\de_{s}^{\rm (g)}]^2 \ran \ne b^2 \lan
\de_{s}^2 \ran$ and the difference for, say, $b = 2$ and $\Omega = 1$
is $\approx 35$\%.

The redshift-space mass density skewness was computed for scale-free
power spectra by Hivon \etal (1995). They found that it depends very
weakly on $\Omega$ and can be well approximated by the real-space
skewness, except in the case that $\Omega$ is close to unity {\em
and\/} the spectral index $n \ga 0$. For more realistic power spectra
the calculation can be done in an analogous way and the result is also
expected to be well approximated by the corresponding real-space
value. Again, it means only that redshift-space mapping globally
changes the third moment and the square of the second by nearly the
same factor.

The redshift-space galaxian density skewness of the {\it IRAS\/}
density field was measured directly by Bouchet \etal (1993). Kim \&
Strauss (1998) pointed out that sparse sampling of the {\it IRAS\/}
galaxies makes estimates of the skewness using the moments method
biased low. However, they proposed a method to measure the skewness by
fitting the Edgeworth expansion to the galaxy count probability
distribution function around its maximum, properly accounting for the
shot noise. Kim \& Strauss showed using mock catalogs that this
estimate of skewness is robust to sparse sampling; they then used this
method to measure the skewness of the {\it IRAS\/} density
field. Hence, we know the value of the galaxian skewness if we are to
perform the {\it IRAS\/}--{\sc potent} comparison, or at least know an
effective way to measure it.

A simple form of equation~(\ref{eq:S_3}) enables us to eliminate $b_2$
from equation~(\ref{app:a_2G_g}) for the coefficient $a_2^{\rm
(g,G)}$ (for Gaussian smoothing). Doing this and using the result in
equation~(\ref{eq:dv2_bias_plain}) yields
\begin{eqnarray}
\left.\left\lan \de_{s}^{\rm (g)} \right\ran \right|_{\te} 
\!\!\! &=& \!\!\!  
- \left(1 + {\textstyle \f{1}{3}} \beta \right) \beta^{-1} \te + 
\nonumber \\
\!\!\! &~& \!\!\!
\left[c_2 + {\textstyle \f{1}{6}} 
\left(S_{3 s}^{\rm (g)} - b^{-1} S_{3 s} \right) 
\right] \beta^{-2} \left(\te^2 - \s_{\te}^2 \right) 
\,,
\label{eq:bias_dvr3}
\end{eqnarray}
where $c_2$ is given by equation~(\ref{app:c_2}). The parameters
$\Omega$ and $b$ appearing in the above relation are {\em not\/}
degenerate: while in the linear coefficient they appear only via their
combination $\beta$, in the quadratic coefficient they clearly do
not. Therefore, measuring the linear coefficient and the parabolic
correction in the galaxian den\-si\-ty versus velocity-divergence
relation one can, at least in principle, measure $\Omega$ and $b$
separately.

\section{Summary and conclusions}
\label{sec:summ}
The main purpose of the present paper is to outline a self-consistent
method of breaking the $\Omega$--bias degeneracy in density--velocity
comparisons by nonlinear effects. A problem which was present in
previous similar attempts (Dekel \etal 1993; see also \cl) is that
they assumed linear biasing. However, if we account for nonlinear
dynamics, in general we should account for nonlinear biasing as
well. There is also another problem which plagues all real-space
comparisons, even these performed at the linear level, aiming merely
at inferring the estimate of $\beta$. Namely, in the process of
reconstruction of the real-space density from the redshift-space
density one has to assume the value of $\beta$, the parameter which is
to be subsequently estimated.\footnote{When attempting to reconstruct
the {\em nonlinear\/} real-space density field, this problem would get
even worse: one would have to assume simultaneously the values of
$\beta$, linear bias and nonlinear (quadratic) bias.}

With nonlinear biasing accounted for in the comparison, the resulting
system of equations for bias parameters and $\Omega$ is
underconstrained. As recently pointed out by B99, an additional
constraint on $\Omega$ and bias can be inferred from the density field
alone (namely, its skewness), so additional types of observations are
unnecessary. In other words, there is enough information in the
density field and the corresponding velocity field to disentangle
$\Omega$ from bias. As for the problem with the real-space density
reconstruction, Nusser \& Davis (1994) proposed to solve it by
performing the comparison directly in redshift space.

One might thus think that to break the $\Omega$--bias degeneracy in
density--velocity comparisons by nonlinear effects one should: (a)
account for nonlinear biasing, (b) add an additional constraint on it
from the density field (e.g., the skewness) and (c) perform the
comparison in redshift space. However, here we face a new problem: the
redshift-space velocity field is rotational at second order, so in the
nonlinear regime it cannot be reconstructed from its radial component
only. Since the method of breaking the degeneracy is based on
nonlinear terms in the density--velocity relation, the redshift-space
analysis is also inappropriate.

A way out of this dilemma is to relate the {\em redshift-space\/}
density field directly to the {\em real-space\/} velocity field. Such
a relation allows one to avoid simultaneously problems with the
reconstruction of the density in real space and of the 
three-dimensional velocity in redshift space. In the present paper I
derived this relation explicitly up to second-order terms. The
relation is local, i.e.\ it expresses density at a given point in
terms of the velocity derivatives at the same point. (Strictly
speaking, it expresses the redshift-space density at a point
$\check{\bfs} = \bfr$ in terms of the real-space velocity derivatives
at $\bfr$.) Therefore, it can be straightforwardly applied to
so-called density--density comparisons and I did this here. It can be
also applied to velocity--velocity comparisons; I will address this
point elsewhere.

Next, I rigorously accounted for the effects of smoothing of the
evolved density and velocity fields. This was necessary because
smoothing changes the degree of nonlinearity in the density--velocity
relation considerably.

Finally, I derived a second-order relation between the redshift-space
{\em galaxian\/} density and the real-space velocity divergence, under
an assumption of nonlinear but local bias. This relation
(eq.~\ref{eq:dv2_bias_plain}) enables one to perform density--velocity
comparisons in a model-independent way. When combined with an
additional measurement of the skewness of the galaxy density field
(eq.~\ref{eq:bias_dvr3}), it also enables in principle to solve for
$\Omega$ and bias separately in density--density comparisons. In
short, the idea is to measure the amplitude of the nonlinear
corrective term in the relation.

The density versus the velocity-divergence relation has a scatter even
at linear order. This scatter is relatively small when compared
to the linear term in equation~(\ref{eq:bias_dvr3}). This means that
the linear estimate of $\beta$ is affected by the scatter in a rather
weak way. In any case, a fairly large number of independent volumes in
the comparison will enable to suppress this scatter, allowing one to
estimate the linear coefficient and hence $\beta$ accurately.

The idea of breaking the $\Omega$--bias degeneracy presented here is
somewhat related to that given by Pen~(1998). In order to measure bias
and $\Omega$ separately, Pen proposed to use simultaneously redshift
distortions of the power spectrum and of the skewness. However, while
in his method the estimate of $\Omega$ is based mostly on the redshift
distortion of the galaxy skewness (which is quite a small effect),
here it is based on the nonlinear effects in the redshift-space
density versus real-space velocity relation. These effects are
considerably stronger than in the corresponding real-space relation.  

Quality of the current velocity data is probably too poor to perform a 
degeneracy-breaking analysis similar to proposed here. The scatter in
the (real-space) density--velocity diagram of the {\it IRAS\/}--{\sc
potent} comparison by Sigad \etal (1988) is huge and it results in
significant errors in the estimate of $\beta$. If even the linear
coefficient is to some extent uncertain we cannot hope to reliably
estimate the nonlinear corrections.

The analysis in its present form is still very simplistic. Points that
should be addressed in further studies in the first place are: (a)
stochastic nature of biasing; (b) higher than second-order effects,
which may play a role for scales smaller than $\simeq 12$ \hmpc; (c)
relaxing the distant-observer assumption, not strictly valid for
current density--velocity comparisons.\footnote{Nevertheless, Kaiser's
classical formula for the redshift-space linear power spectrum,
derived in the distant-observer approximation, is commonly used in
analyses of redshift distortions in galaxy surveys like {\it IRAS}.}
Recently, Taruya \& Soda (1998) have derived a mildly nonlinear galaxy
density--mass density relation in a stochastic bias model. The
calculation of the galaxy density--velocity divergence relation in
such a model is mathematically entirely analogous. Higher-orders
effects are complex, but tractable; here I computed them for
unsmoothed fields. Similarly, calculation of the density--velocity
relation without the simplifying assumption of a distant observer,
though more complex, is certainly possible.

The purpose of this paper was to perform a {\em first step\/} towards
self-consistent breaking the $\Omega$-bias degeneracy in
density--velocity comparisons by nonlinear effects. I showed that the
effects in the comparisons proposed here are fairly strong and that
the fact that they are of the same order as possible nonlinearities in
the biasing relation is not a fundamental obstacle. The method of
breaking the $\Omega$--bias degeneracy presented here can be worked
out in full detail when data of sufficient quality exists. I mean
primarily catalogs of less noisy, more densely and homogeneously
sampled peculiar velocities.

\section*{Acknowledgments}
I thank Adi Nusser for the idea of studying the density--velocity
relation in redshift space, Stephane Colombi for useful technical
discussions and Michael Blanton, Rom\'an Scoccimarro and Radek Stompor
for valuable comments. I am grateful to Bohdan Paczy\'nski for his
hospitality at Princeton where part of this work was done. I am also
grateful to an anonymous referee for pointing out an error in an
earlier version of this paper. This research has been supported in
part by the Polish State Committee for Scientific Research grants
No.~2.P03D.008.13 and 2.P03D.004.13.

\onecolumn

\appendix
\section{Derivation of equation~(\lowercase{\ref{eq:scocci}})}
\label{app:scocci}

In this appendix I present the derivation of exact
relation~(\ref{eq:scocci}), expressing the redshift-space density at
$\check{\bfs} = \bfr$ in terms of the real-space density and velocity
at $\bfr$. In the distant-observer approximation, the line-of-sight is
taken as a fixed direction, which will be denoted here by $\hat{z}
\equiv \hat{r_3}$. The real-to-redshift mapping~(\ref{eq:realtored})
reduces then to $\bfs = \bfr - f \tu(\bfr) \hat{z}$, where $\tu$ is
the $z$-component of the scaled velocity field, defined by
equation~(\ref{eq:scaled_vel}). From the conservation of the number of
galaxies we have $1 + \de_{s}(\bfs) = J(\bfr)[1 + \de(\bfr)]$, where
$J(\bfr)$ is the Jacobian of the mapping. In the plane-parallel
approximation, $J(\bfr) = 1 - f \tu'(\bfr)$ (exactly), where $' \equiv
\partial/\partial z$. This yields

\be
\de_{s}(\bfs) = \f{\de(\bfr) + 1 - J(\bfr)}{J(\bfr)} 
= \f{\de(\bfr) + f \tu'(\bfr)}{1 - f \tu'(\bfr)} 
\label{eq:ds-dr_distant}
\ee 
(eq.~3 of Scoccimarro \etal 1998). Fourier transforming the above
equation we obtain
\be 
\de_{s}(\bfk) \equiv \int \f{{\rm d}^3 s}{(2 \pi)^3} {\rm e}^{-i
\bfk \cdot \bfs} \de_{s}(\bfs) = \int \f{{\rm d}^3 r}{(2 \pi)^3} {\rm
e}^{-i \bfk \cdot \bfr} {\rm e}^{i f k_z \tu(\bfr)} \left[\de(\bfr) +
f \tu'(\bfr)\right] 
\label{eq:ds(k)}
\ee
(eq.~4 of Scoccimarro \etal 1998). Thus we have expressed the Fourier
transform of the density contrast in redshift space as a real-space
integral involving the real-space density and velocity fields. The
inverse Fourier transform of equation~(\ref{eq:ds(k)}) is

\be \de_{s}(\bfs) \equiv \int {\rm d}^3 k \, {\rm e}^{i \bfk \cdot \bfs}
\de_{s}(\bfk) = \int \f{{\rm d}^3 r \, {\rm d}^3 k}{(2 \pi)^3} 
\left[\de(\bfr) + f \tu'(\bfr)\right]
{\rm e}^{i f k_z \tu(\bfr)} {\rm e}^{i \bfk \cdot (\bfs - \bfr)}
 \,,
\label{eq:ds(s)}
\ee
where I have changed the order of integration. Expanding the first
exponent of the integrand we have 

\be {\rm e}^{i f k_z \tu(\bfr)} {\rm e}^{i \bfk \cdot (\bfs - \bfr)} =
\sum_{n=0}^\infty \f{[f \tu(\bfr)]^n}{n!} (i k_z)^n {\rm e}^{i \bfk \cdot
(\bfs - \bfr)} =
\sum_{n=0}^\infty \f{(-1)^n f^n \tu^n(\bfr)}{n!} \f{\partial^n}{\partial
z^n} {\rm e}^{i \bfk \cdot
(\bfs - \bfr)} 
\,,
\label{eq:exp}
\ee
hence

\be 
\de_{s}(\bfs) = 
\sum_{n=0}^\infty \f{(-1)^n f^n}{n!} \int {\rm d}^3 r \, \tu^n(\bfr) 
\left[\de(\bfr) + f \tu'(\bfr)\right] \f{\partial^n}{\partial z^n}
\int \f{{\rm d}^3 k}{(2 \pi)^3} {\rm e}^{i \bfk \cdot (\bfs - \bfr)} 
\,. 
\label{eq:ds(s)2}
\ee
The integral over $k$ yields the Dirac delta distribution, $\de_{\rm
D}(\bfs - \bfr)$. Integrating the remaining integral over $r$ by parts
we obtain

\be
\de_{s}(\bfs) = \sum_{n = 0}^{\infty} \f{1}{n!} f^n
\left[\tu_{}^n (\de + f \tu_{}')\right]^{(n)} \bigr|_{\check{\bfr} =
\bfs} \,,
\label{eq:ds_final}
\ee 
where the superscript `$^{(n)}$' denotes `$\partial^n/\partial
z^n$'. Thus, if we select a point $\bfs$ in redshift space, to
estimate $\de_{s}(\bfs)$ we evaluate the expression on the RHS of the
above equation at a real-space point $\check{\bfr} =
\bfs$. Conversely, if we first fix a point $\bfr$ in real space,
evaluating the RHS of the above equation at $\bfr$ we estimate the
redshift-space density contrast at a redshift-space point
$\check{\bfs} = \bfr$. This yields equation~(\ref{eq:scocci}).

\section{Calculation of the coefficient $\bmath{\lowercase{a_2}}$}
\subsection{The case of no bias}
\label{app:nobias}

Our purpose here is to compute the coefficient $a_2$, given by
equation~(\ref{eq:a_2}), for smoothed evolved fields. The calculation
is much easier when performed in Fourier space. Denote the Fourier
transform of the linear real-space density field $\de_1$ by
$\eps(\bfk)$. According to linear theory $\tte_1 = \de_1$, so
$\eps(\bfk)$ is also the Fourier transform of the linear scaled
velocity divergence. We have

\be
\left\lan \tte_1^2 \right\ran = \s^2(R)
\,,
\label{app:var}
\ee
where $\s^2(R)$ is the linear variance of the real-space density field
smoothed over a scale $R$. To compute the quantity $\left\lan \tte_1^2
\Delta_2 \right\ran$, we need the Fourier transform of $\Delta_2$
(defined by eq.~\ref{eq:Delta}). Following Heavens \etal (1998)
I find that

\be
\Delta_{2}(\bfk) = \int \f{{\rm d}^3 p\, {\rm d}^3 q}{(2 \pi)^2} 
\de_D(\bfk - \bfp - \bfq) T_2(\bfp,\bfq) \eps(\bfp)
\eps(\bfq)
\,,
\label{app:Del2_Fourier}
\ee
where 
\be
T_2(\bfp,\bfq) =  
\f{2}{7} \left(1 - \cos^2{\varte}\right)
+ \f{f}{2} \left[\mu^2_{p} + \mu^2_{q} + \mu_{p} \mu_{q} 
\left(\f{p}{q} + \f{q}{p}\right) \right] + 
f^2 \left[\mu^2_{p} \mu^2_{q} + \f{\mu_{p} \mu_{q}}{2} 
\left(\mu^2_{p} \f{p}{q} + \mu^2_{q} \f{q}{p}\right) \right]
.
\label{app:T_2}
\ee

In the above, $\cos{\varte} = \hat{\bfp} \cdot \hat{\bfq}$ and for any
vector $\hat{\bfk} \equiv \bfk / k$. The remaining quantities are
$\mu_{p} = \hat{\bfp} \cdot \hat{\bfs}$ and $\mu_{q} = \hat{\bfq}
\cdot \hat{\bfs}$, where $\hat{\bfs}$ is a unit vector from the
observer to the galaxy; in the distant-observer approximation one
assumes $\hat{\bfs}$ is constant within the smoothing radius.

For Gaussian initial fluctuations, straightforward calculation yields

\be
{\textstyle \f{1}{2}} \left\lan \tte_1^2 \Delta_2 \right\ran = 
\int \f{{\rm d}^3 p\, {\rm d}^3 q}{(2 \pi)^6}
W(p) W(q) W(|\bfp + \bfq|) P(p) P(q) T_2(\bfp,\bfq) 
\label{app:mom}
\ee 
(see e.g.\ Juszkiewicz, Bouchet \& Colombi 1993). Here, $P$ is the
real-space linear mass power spectrum and $W$ is the Fourier transform
of the window function. The calculation of the above integral with the
kernel $T_2$ given by equation~(\ref{app:T_2}) is similar to the
calculation of the skewness of the redshift-space density field.
Following Hivon \etal (1995) I note that the angular integration of
the above expression for a fixed $\varte$ is, up to a multiplicative
factor, equivalent to averaging over different orientations of
$\bfs$. We then have $\bfp = (0,0,1) p$, $\bfq =
(\sin\varte,0,\cos\varte) q$ and $\bfs = (\cos\phi'
\sin\varte',\sin\phi' \sin\varte',\cos\varte') s$, so

\be
\mu_p = \cos\varte'
\,,
\label{app:mu_p}
\ee 
\be
\mu_q = \cos\phi' \sin\varte' \sin\varte + \cos\varte' \cos\varte 
\,,
\label{app:mu_q}
\ee 
and the averaging over the polynomials of $\mu_p$ and $\mu_q$ given
$\varte$ is effectively over the trigonometric functions of $\phi'$
and $\varte'$. The result is

\be
\lan T_2(\bfp,\bfq) \ran_{\mu_p \mu_q}|_\varte =
\f{2}{7} + \f{f}{3} + \f{f^2}{15} 
+ \left(\f{f}{6} + \f{f^2}{10} \right) 
\cos\varte \left(\f{p}{q} + \f{q}{p} \right)
- \left(\f{2}{7} - \f{2 f^2}{15} \right) \cos^2\varte
\,.
\label{app:T_2mean}
\ee

Using equations~(\ref{app:var}) and~(\ref{app:mom}) in
equation~(\ref{eq:a_2}) yields
 
\be
a_2 = \int \f{{\rm d}^3 p \, {\rm d}^3 q}{(2 \pi)^6 \s^4} 
W(p) W(q) W(|\bfp + \bfq|) P(p) P(q) \lan T_2(\bfp,\bfq) \ran
\,,
\label{app:a_2int}
\ee 
where $\lan T_2 \ran$ is given by expression~(\ref{app:T_2mean}). The
calculation of the coefficient $a_2$ is now entirely analogous to the
calculation of the skewness of the {\em real-space\/} density
field. This is so because the averaged kernel has a similar functional
dependence on $\varte$, $p$ and $q$ to that of the second-order
solution for the real-space density field (see e.g.\ Juszkiewicz \etal
1993), with different coefficients. In the case of unsmoothed
fields $W(k) \equiv 1$ and we obtain

\be
a_2^{(\rm un)} = \f{4}{21} + \f{f}{3} + \f{f^2}{9}
\,.
\label{app:a_2unsmooth}
\ee

Comparing~(\ref{app:a_2unsmooth}) with expressions~(\ref{eq:tophat})
and~(\ref{eq:mean_dv3}), we see that $a_2^{(\rm un)}$ is equal to the
value predicted by the spherical collapse model.

To study the effects of smoothing I will consider power spectra with a
power-law form

\be    
P(k) = C k^n, \ \ -3\leq n \leq 1
\,,
\label{app:ps}
\ee
where $C$ is a normalization constant. For more realistic spectra, the
value of $a_2$ can be very well approximated by the result for
scale-free spectra with the effective index at a smoothing scale
defined as

\be   
n_{\rm eff} = - \frac{R}{\sigma^2} \frac{d \sigma^{2}(R)}{d R} - 3
\label{app:n_eff}
\ee 
(\cl). For the case of a spherical top-hat smoothing function,
using the results of Bern\-ard\-eau (1994) I find that

\be
a_2^{(\rm TH)} = a_2^{(\rm un)} 
- (n+3) \left( \f{f}{18} + \f{f^2}{30} \right)
\,.
\label{app:a_2TH}
\ee

For the most interesting case of Gaussian smoothing, using 
the results of {\L}okas \etal (1995) I obtain

\be
a_2^{(\rm G)} = \f{4}{21} F(5/2) 
+ \f{f}{6} \left[ F(3/2) - \f{n}{3} F(5/2) \right]
+ \f{f^2}{10} \left[ F(3/2) - \f{3n + 8}{9} F(5/2) \right]
\,.
\label{app:a_2G}
\ee
Here, 

\be
F(3/2) \equiv \ _{2} F_{1} \left( \f{n+3}{2}, \f{n+3}{2},
      \f{3}{2}, \f{1}{4} \right)
\,,
\label{app:F32}
\ee
\be
F(5/2) \equiv \ _{2} F_{1} \left( \f{n+3}{2}, \f{n+3}{2},
      \f{5}{2}, \f{1}{4} \right)
\label{app:F52}
\ee 
and $_{2} F_{1}$ is the hypergeometric function. The
hypergeometric function can be expanded in a series of powers of $n+3$
which converges rapidly. Fitting the expansion by a polynomial in
$n+3$ in the range $-3 \le n \le 1$ yields low-order accurate
fits. Namely,

\be
F(r) = 1 + d_2^{(r)} (n+3)^2 + d_3^{(r)} (n+3)^3
\label{app:Ffit}
\,,
\ee
where 

\be
d_2^{(3/2)} = 0.03239, \qquad d_3^{(3/2)} = 0.009364
\label{app:d32}
\ee
and

\be
d_2^{(5/2)} = 0.02183, \qquad d_3^{(5/2)} = 0.003531
\label{app:d52}
\,.
\ee
The fit to the function $F(3/2)$ is accurate to $0.6 \%$ or less in
the range $-3 \le n \le 1$; the fit to $F(5/2)$ is accurate to $0.2
\%$ or less.

As in the case of top-hat smoothing, for $n = -3$ we have $a_2^{(\rm
G)} = a_2^{(\rm un)}$ and for other values of the spectral index
$a_2^{(\rm G)} \ne a_2^{(\rm un)}$. Thus, only for $n = -3$ are the
values of the coefficient $a_2$ for smoothed fields equal to that
given by the spherical collapse model. 

Finally, we recover the real-space values of $a_2$ (\cl, B99) by
setting $f=0$ in equations~(\ref{app:a_2unsmooth}), (\ref{app:a_2TH})
and~(\ref{app:a_2G}):

\be
a_{2,{\rm real}}^{(\rm TH)} = \f{4}{21} 
\label{app:a_2THreal}
\ee
and
\be
a_{2,{\rm real}}^{(\rm G)} = \f{4}{21} F(5/2) 
\,.
\label{app:a_2Greal}
\ee

\subsection{The case of bias}
\label{app:bias} 
Our purpose here is to compute the coefficient $a_2^{\rm (g)}$, entering
in the relation~(\ref{eq:dv2_bias}) between the redshift-space {\em
galaxian\/} density and the real-space velocity divergence. All
calculations are similar to those in the case of no bias, presented in
the previous Subsection. Using equation~(\ref{eq:dv_bias2}),
analogously to equations~(\ref{eq:mean_dv})--(\ref{eq:Delta}) we can
write

\be
\lan \de_{s}^{\rm (g)} \ran \vert_{\tte} = 
b \left(1 + {\textstyle \f{1}{3}} \beta \right) \tte
+ \lan \Delta_2^{\rm (g)} \ran|_{\tte}
\,,
\label{eq:mean_dv_bias}
\ee
where

\be
\Delta_2^{\rm (g)} \equiv 
f \left[\tu(b \tte + f \tu')\right]'
+ {\textstyle \f{4}{21}} b 
\left(\tte^2 - {\textstyle \f{3}{2}} \tS^2\right) 
+ {\textstyle \f{1}{2}} b_2
\left(\tte^2 - \lan \tte^2 \ran \right)
\,.
\label{eq:Delta_g}
\ee
The coefficient $a_2^{\rm (g)}$ is given by the following equation
(cf.~eq.~\ref{eq:a_2}):

\be
a_2^{\rm (g)} = {\textstyle \f{1}{2}} \left\lan \tte_1^2 \right\ran^{-2} 
\left\lan \tte_1^2 \Delta_2^{\rm (g)} \right\ran  
\,.
\label{eq:a_2_g}
\ee
The Fourier transform of $\Delta_2^{\rm (g)}$ is 

\be
\Delta_{2}^{\rm (g)}(\bfk) = \int \f{{\rm d}^3 p\, {\rm d}^3 q}{(2 \pi)^2} 
\de_D(\bfk - \bfp - \bfq) T_2(\bfp,\bfq) \eps(\bfp)
\eps(\bfq) \, - \, 4 \pi^3 b_2 \s^2 \de_D(\bfk)
\,,
\label{app:Del2_Fourier_g}
\ee
where already an averaged form of the kernel is 
\be
\lan T_2^{\rm (g)}(\bfp,\bfq) \ran_{\mu_p \mu_q}|_\varte =
\f{2 b}{7} + \f{3 b_2}{4} + \f{b f}{3} + \f{f^2}{15} 
+ \left(\f{b f}{6} + \f{f^2}{10} \right) 
\cos\varte \left(\f{p}{q} + \f{q}{p} \right)
- \left(\f{2 b}{7} - \f{2 f^2}{15} \right) \cos^2\varte
\,.
\label{app:T_2mean_g}
\ee
In the expression~(\ref{app:Del2_Fourier_g}), the last term is the Fourier
transform of the term $(b_2/2) \lan \tte^2 \ran$. 

Using equations~(\ref{eq:a_2_g}) and~(\ref{app:Del2_Fourier_g}) yields
(cf. eq.~\ref{app:a_2int})

\be
a_2^{\rm (g)} = \int \f{{\rm d}^3 p \, {\rm d}^3 q}{(2 \pi)^6 \s^4} 
W(p) W(q) W(|\bfp + \bfq|) P(p) P(q) \lan T_2^{\rm (g)}(\bfp,\bfq) \ran 
- \f{b_2}{4} \,.
\label{app:a_2int_g}
\ee
For Gaussian-smoothed fields, the result of the integration is 

\be
a_2^{\rm (g,G)} = \f{b_2}{2} + b^2 c_2 \,,
\label{app:a_2G_g}
\ee
where

\be 
c_2 = \f{4\, b^{-1}}{21} F(5/2) 
+ \f{\beta}{6} \left[ F(3/2) - \f{n}{3} F(5/2) \right]
+ \f{\beta^2}{10} \left[ F(3/2) - \f{3n + 8}{9} F(5/2) \right] ,
\label{app:c_2}
\ee 
$\beta \equiv f(\Omega)/b$ and $F(3/2)$ and $F(5/2)$ are defined in
equations~(\ref{app:F32})--(\ref{app:F52}). In the limit of no bias
($b = 1$ and $b_2 = 0$), equations~(\ref{app:a_2G_g})--(\ref{app:c_2})
reduce to equation~(\ref{app:a_2G}), as expected. The result for a
top-hat smoothing can be obtained by replacing the factors $F(3/2)$
and $F(5/2)$ by unity. The result for unsmoothed fields then follows
from setting the spectral index equal to $ -3$.

\section{Gaussian conditional averages}
\label{app:u'}
Our purpose in this Section is to compute the quantity $\lan \tu'^2
\ran|_{\tte}$ at leading, i.e.\ linear, order. We may assume therefore
that $\tu'$ and $\tte$ are Gaussian variables. For such variables,

\be
\lan \tu' \ran|_{\tte} = r \f{\s_{\tu'}}{\s_{\tte}} \tte
\label{appb:mean_in}
\,.
\ee

Here, $\s_{\tu'}^2$ ($\s_{\tte}^2$) is the variance of the variable
$\tu'$ ($\tte$) and $r$ is the correlation coefficient defined as

\be
r = \f{\lan \tu' \tte \ran}{\s_{\tu'} \s_{\tte}}
\label{appb:r}
\,.
\ee

Using results of Appendix~\ref{app:nobias} we have $\s_{\tu'}^2 = (1/5)
\s_{\tte}^2$ and $r = \sqrt{5}/3$. This yields

\be
\lan \tu' \ran|_{\tte} = {\textstyle \f{1}{3}} \tte
\label{appb:mean_fin}
\,,
\ee
a result which is obvious by isotropy. The conditional
variance is 

\be
\lan \tu'^2 \ran|_{\tte} - \lan \tu' \ran|_{\tte}^2 = 
\left(1 - r^2 \right) \s_{\tu'}^2 =
{\textstyle \f{4}{45}}\s_{\tte}^2
\label{appb:var}
\,,
\ee 
hence finally 

\be
\lan \tu'^2 \ran|_{\tte} = 
{\textstyle \f{1}{9}} \tte^2 + {\textstyle \f{4}{45}}\s_{\tte}^2 
\label{appb:fin}
\,.
\ee


\begin{thebibliography}{}
  \bibitem[\protect\citename{Bardeen \etal }1986]{bar86} 
  Bardeen J., Bond J. R., Kaiser N., Szalay A., 1986, ApJ, 304, 15
\bibitem[\protect\citename{Bernardeau }1994]{ber94} 
  Bernardeau F., 1994, ApJ, 433, 1 
\bibitem[\protect\citename{Bernardeau \etal }1999]{ber99} 
  Bernardeau F., Chodorowski M. J., {\L}okas E. L., Stompor R., 
  Kudlicki A., 1999, preprint astro-ph/9901057 (B99)
\bibitem[\protect\citename{Blanton \etal }1998]{blanton} 
  Blanton M., Cen R., Ostriker J. P., Strauss M. A., 
  1998, preprint astro-ph/9807029 
\bibitem[\protect\citename{Bouchet \etal }1993]{bou93} 
  Bouchet F. R., Strauss M. A., Davis M., Fisher K. B., Yahil A.,
  Huchra J. P., 1993, ApJ, 417, 36
\bibitem[\protect\citename{Catelan \etal }1995]{cat95} 
  Catelan P., Lucchin F., Matarrese S., Moscardini L.,
  1995, MNRAS, 276, 39 
\bibitem[\protect\citename{Cen \& Ostriker }1992]{co92} 
  Cen R., Ostriker J. P., 1992, ApJ, 399, L113
\bibitem[\protect\citename{Chodorowski }1990]{ch90}
  Chodorowski M. J., 1990, Acta Astr., 40, 11  
\bibitem[\protect\citename{Chodorowski }1997]{ch97}
  Chodorowski M. J., 1997, MNRAS, 292, 695
\bibitem[\protect\citename{Chodorowski \& {\L}okas }1997]{cl97}
  Chodorowski M. J., {\L}okas E. L., 1997, MNRAS, 287, 591 (\cl)
\bibitem[\protect\citename{Chodorowski \etal }1998]{chod98}
  Chodorowski M. J., {\L}okas E. L., Pollo A., Nusser A., 1998, MNRAS,
  300, 1027 
\bibitem[\protect\citename{Davis \& Geller }1976]{dg76}
  Davis M., Geller M. J., 1976, ApJ, 208, 13 
\bibitem[\protect\citename{Davis \etal }1985]{dav85}
  Davis M., Efstathiou G., Frenk C. S., White S. D. M., 1985, ApJ,
  292, 371 
\bibitem[\protect\citename{Dekel \& Silk }1986]{ds86}
  Dekel A., Silk J., 1986, ApJ, 303, 39 
\bibitem[\protect\citename{Dekel \& Lahav }1998]{dl98}
  Dekel A., Lahav O., 1998, preprint astro-ph/9806193
\bibitem[\protect\citename{Dekel \etal }1993]{dbysd}
  Dekel A., Bertschinger E., Yahil A., Strauss M., Davis M.,
  Huchra J., 1993, ApJ, 412, 1
\bibitem[\protect\citename{Dekel \etal }1998]{dek98}
  Dekel A. et al., 1998, preprint astro-ph/9812197
\bibitem[\protect\citename{Dressler }1980]{dress80}
  Dressler A., 1980, ApJ, 236, 351
\bibitem[\protect\citename{Fisher \etal }1995]{fish95}
  Fisher K. B., Lahav O., Hoffman Y., Lynden-Bell D., Zaroubi S.,
  1995, MNRAS, 272, 885
\bibitem[\protect\citename{Fry }1994]{fry94}
  Fry J. N., 1994, Phys.\ Rev.\ Lett., 73, 215 
\bibitem[\protect\citename{Fry \& Gazta\~naga }1993]{fg93}
  Fry J. N., Gazta\~naga E., 1993, ApJ, 413, 447
\bibitem[\protect\citename{Ganon \etal }1999]{gan99}
  Ganon G., Dekel A., Mancinelli P. J., Yahil A., 1999,
  in preparation
\bibitem[\protect\citename{Giavalisco \etal }1998]{gia98} 
  Giavalisco M. \etal, 1998, preprint astro-ph/9802318
\bibitem[\protect\citename{Giovanelli, Haynes \& Chincarini }1986]{ghc86}
  Giovanelli R., Haynes M. P., Chincarini G. L., 1986, ApJ, 300, 77
\bibitem[\protect\citename{Gramann }1993]{gramann}
  Gramann M., 1993, ApJ, 405, L47
\bibitem[\protect\citename{Guzzo \etal }1997]{guz97}
  Guzzo L., Strauss M. A., Fisher K. B., Giovanelli R., Haynes M. P.,
  1997, ApJ, 489, 37
\bibitem[\protect\citename{Hamilton }1997]{ham97}
  Hamilton A. J. S., 1997, in {\em Ringberg Workshop on Large-Scale
  Structure,\/} ed.\ D. Hamilton, Kluwer Academic, preprint 
  astro-ph/9708102 
\bibitem[\protect\citename{Heavens, Matarrese \& Verde }1998]{hmv98}
  Heavens A. F., Matarrese S., Verde L., 1998, preprint
astro-ph/9808016 
\bibitem[\protect\citename{Hermit \etal }1996]{herm96} 
  Hermit S., Santiago B. X., Lahav O., Strauss M. A., Davis M., 
     Dressler A., Huchra J. P., 1996, MNRAS, 283, 709
\bibitem[\protect\citename{Hivon \etal }1995]{hetal95} 
  Hivon E., Bouchet F. R., Colombi S., Juszkiewicz R., 1995, A\&A, 298, 643
\bibitem[\protect\citename{Juszkiewicz, Bouchet \& Colombi }1993]{jbc93}
  Juszkiewicz R., Bouchet F. R., Colombi S., 1993, ApJ, 412, L9
\bibitem[\protect\citename{Juszkiewicz \etal }1995]{jusz}
  Juszkiewicz R., Weinberg D. H., Amsterdamski P., Chodorowski M. J.,
  Bouchet F. R., 1995, ApJ, 442, 39 
\bibitem[\protect\citename{Kaiser }1984]{kai84}
  Kaiser N., 1984, ApJL, 284, L9
\bibitem[\protect\citename{Kaiser }1987]{kai87}
  Kaiser N., 1987, MNRAS, 227, 1 
\bibitem[\protect\citename{Kauffman, Nusser \& Steinmetz }1997]{kns}
  Kauffman G., Nusser A., Steinmetz M., 1997, MNRAS, 286, 795
\bibitem[\protect\citename{Kim \& Strauss }1998]{ks98}
  Kim R. S., Strauss M. A., 1998, ApJ, 493, 39
\bibitem[\protect\citename{Loveday \etal }1996]{lov96}
  Loveday J., Efstathiou G., Maddox S. J., Peterson B. A., 1996, ApJ,
  468, 1
\bibitem[\protect\citename{{\L}okas \etal }1995]{lok95}
  {\L}okas E. L., Juszkiewicz R., Weinberg D. H., Bouchet F. R., 
  1995, MNRAS, 274, 730
\bibitem[\protect\citename{Mancinelli \& Yahil }1995]{my95}
  Mancinelli P. J., Yahil A., 1995, ApJ, 452, 75 
\bibitem[\protect\citename{Nusser \& Davis }1994]{nd94} 
  Nusser A., Davis M., 1994, ApJ, 421, L1
\bibitem[\protect\citename{Pen }1998]{pen} 
  Pen U., 1998, ApJ, 504, 601
\bibitem[\protect\citename{Santiago \& Strauss }1992]{ss92} 
  Santiago B. X., Strauss M. A., 1992, ApJ, 387, 9
\bibitem[\protect\citename{Scoccimarro, Couchman \& Frieman }1998]{scf92}
  Scoccimarro R., Couchman H. M. P., Frieman J. A., 1998, 
  preprint astro-ph/9808305 
\bibitem[\protect\citename{Sigad \etal }1998]{sig98}
  Sigad Y., Eldar A., Dekel A., Strauss M. A., Yahil A., 
  1998, ApJ, 495, 516  
\bibitem[\protect\citename{Strauss \etal }1992]{st92}
  Strauss M. A., Yahil A., Davis M., Huchra J. P., Fisher K., 1992,
  ApJ, 397, 395
\bibitem[\protect\citename{Taruya \& Soda }1998]{ts98}
  Taruya A., Soda J., 1998, preprint astro-ph/9809204
\bibitem[\protect\citename{Taylor \& Hamilton }1996]{th96}
  Taylor A. N., Hamilton A. J. S., 1996, MNRAS, 282, 767
\bibitem[\protect\citename{Taylor \& Valentine }1999]{tv99}
  Taylor A. N., Valentine H., 1999, preprint astro-ph/9901171 
\bibitem[\protect\citename{Tegmark \& Bromley }1998]{tb98}
  Tegmark M., Bromley B. C., 1998, preprint astro-ph/9809324 
\bibitem[\protect\citename{Tegmark \& Peebles }1998]{tp98}
  Tegmark M., Peebles P. J. E., 1998, ApJ, 500, L79
\bibitem[\protect\citename{Yahil \etal }1991]{ya91}
  Yahil A., Strauss, M. A., Davis, M., Huchra J. P., 1991, ApJ, 372,
  380 

\end{thebibliography}
\end{document}